\documentclass{aa}
\batchmode

\usepackage{geometry} 
\usepackage{float} 
\usepackage{amsmath, amssymb}
\usepackage{graphicx} 

\begin{document} 

\title{Kozai-driven mass loss of the circumbinary disk in D9 in orbit around the supermassive black hole Sgr A*}
\titlerunning{Kozai-driven mass loss in the D9 disk}

   \author{
        Yannick Badoux\inst{1}
        \and
        Lucas Pouw\inst{1}
        \and
        Tim van der Vuurst\inst{1}
        \and
        Simon Portegies Zwart\inst{1}
        }

   \institute{Leiden Observatory,  Leiden  University,
              PO  Box  9513,  2300  RA, Leiden, The Netherlands \\
              \email{spz@strw.leidenuniv.nl}}

\abstract {The supermassive black hole (Sgr A*) in the Galactic center
  is surrounded by the S-star cluster consisting of young stars on
  eccentric orbits. Recently, the an S-star binary system, called D9,
  was found. Its detection was based on periodic emission in
  Brackett$\gamma$ (Br$\gamma$). Since Br$\gamma$ is considered a
  signature of accretion, this emission could originate from the
  interaction of a binary and its circumbinary disk. However, due to
  the gravitational interaction between Sgr A* and the binary, the
  disk could be short-lived.}
  {We investigated the evolution of the disk around a stellar binary
    while orbiting Sgr A*.}
  {We used the \texttt{AMUSE} framework for coupling a gravity solver
    (for the binary and Sgr. A*) with a hydrodynamics solver (for the
    disk).}
  {We find that, irrespective of the initial disk inner and outer
    radii, it eventually settles between 5.2$a_{\rm in}$ and 0.28 Hill
    radii of the binary. Here, $a_{\rm in}$ is the semimajor axis of
    the binary D9. The inclination of the circumbinary disk follows
    that of the binary, which evolves due to the von
    Zeipel-Lidov-Kozai (vZLK) mechanism induced by Sgr A*. The mean
    eccentricity of the disk is approximately in antiphase with the
    eccentricity evolution of the binary. We find a vZLK timescale of
    $T_\text{vZLK}\approx62.5\,$kyr, which is two orders of magnitude
    shorter than the value reported earlier. As a
    consequence, D9 has undergone multiple vZLK oscillations in its
    lifetime of 2.7 Myr. We find that the disk shows periodic bursts of
    mass loss on the vZLK timescale, suggesting that the mass loss
    itself is in part driven by the vZLK mechanism.}
  {The secular evolution observed in both the binary and the disk are
    consistent with theoretical predictions. We find the disk loses
    $\sim$7\% $\pm$ 2\% of its mass every vZLK cycle. If we
    extrapolate this mass loss, the disk will have 1\% of its current
    mass left after another $\sim$4 Myr. D9 will then be $\sim$6.7 Myr
    old, which is on the same order as the current average age of S
    cluster members. The vZLK-driven mass loss could therefore
    explain the absence of Br$\gamma$ emission from other S cluster
    members.  }

\date{Received <set by editor>, 2025 / Accepted <set by editor>, 2025}
\keywords{Accretion, accretion disks - Galaxy: center - Stars: kinematics and dynamics - Celestial mechanics}
\maketitle

%
\section{Introduction}\label{sect:introduction}

The central parsec of the Galactic center is a dynamical environment, with at its core the supermassive black hole (SMBH) Sagittarius (Sgr) A* \citep{Eckart_1996,Ghez_1998}. Sgr A* is surrounded by the S-star cluster that contains high-velocity stars on elliptical orbits \citep{Ghez_2008,Gillessen_2009}.

Recently, \citet{Peissker_2024} reported the detection of an S-star binary system, called D9. The primary of D9 is characterized as a Herbig Ae star and is therefore expected to have a gaseous circumstellar disk \citep{Mannings_2000}. The secondary would be a T-Tauri star. The binary nature of D9 was hypothesized because of the periodicity in its Brackett$\gamma$ (Br$\gamma$) emission, which is associated with accretion processes of Herbig Ae/Be and T-Tauri stars \citep{Muzerolle_1998,Grant_2022}. 

\citet{Peissker_2024} offer three pathways to produce periodically varying Br$\gamma$ emission in a binary system: \textit{(i)} The secondary disturbs the circumprimary disk. \textit{(ii)} There is a circumbinary disk from which gas periodically accretes onto the binary. \textit{(iii)} The winds of the secondary star interact with a circumprimary or circumbinary disk, causing varying resonance-intercombination lines to shift \citep{Friedjung_2010}.

The circumprimary or circumbinary nature of the detected disk in D9 is uncertain, because a circumbinary disk may obscure the circumprimary disk. However, a hint may be offered by the inferred low disk mass of $M_{\rm disk} \simeq 1.61 \cdot 10^{-6} \, M_{\odot}$. \citet{Peissker_2024} speculate that the low mass may be explained by a possible formation scenario of the system, in which a molecular cloud migrates toward Sgr A* and, during this, D9 forms with a circumbinary disk. During the $2.7^{+1.9}_{-0.3}\,$Myr lifetime of D9, this disk photoevaporates due to stellar winds, until it has the present-day low mass.

\citet{Peissker_2024} also discuss the future of D9. They argue that SMBH-induced von Zeipel-Lidov-Kozai (vZLK) oscillations \citep{Zeipel_1910,Lidov_1962,Kozai_1962,Naoz_2016} with a period comparable to the age of D9 will cause the system to merge in the near future \citep{Stephan_2016}.

\citet{Peissker_2024} argues that D9 is observed at a special moment
in time, just before the binary merges and before the circumstellar
disk is fully evaporated.  Furthermore, the circumbinary disk is
subject to both the gravitational potential of Sgr A* and the stellar
binary, possibly disrupting the disk. Altogether, this makes the
discovery of D9 an improbable event.

We investigate the possibility of a circumbinary disk surviving in the gravitational potential of Sgr A* in the most optimistic scenario, where only secular evolution influences the system. We show that the disk can remain stable for a long time. We also find that the vZLK effect will not cause a merger of the binary. Overall, this makes the discovery of D9 more probable.

\section{Methods}\label{sect:methods}
\subsection{Computational setup}\label{subsect:integrators}
The dynamics of the binary system orbiting the SMBH is handled by the
gravitational N-body code \texttt{Hermite} \citep{Makino_1991}. To
simulate the dynamics of the disk, we used the smoothed-particle
hydrodynamics (SPH) code \texttt{Fi} \citep{Hernquist_Katz_1989,Gerritsen_Icke_1997,Pelupessy_2004}.
The \texttt{Fi} timestep was set to $\sim\,$4 days, which is 1\% of
the binary period.
The gravito-hydrodynamic simulation was implemented with the
Astronomical Multipurpose Software Environment (\texttt{AMUSE};
\citealt{Portegies_Zwart_2009,Portegies_Zwart_2013,Pelupessy_2013,Portegies_Zwart_2026_book}). The gravity and hydrodynamics solvers
were coupled with a bridge \citep{Fujii_2007}. We used the classic
second-order bridge integration scheme in \texttt{AMUSE}. A
higher-order bridge is not needed in our case, as \texttt{Fi} is a
second-order integrator. The bridge timestep was set to 10 times the
hydrodynamics solver timestep ($\sim$ 40 days).

We set up particle sinks on both stars and Sgr A* that can catch infalling SPH particles. We find no accretion onto either the stars or Sgr A*.

\subsection{Initial conditions}\label{subsect:initialcond}
\begin{figure}
    \centering
    \includegraphics[width=\linewidth]{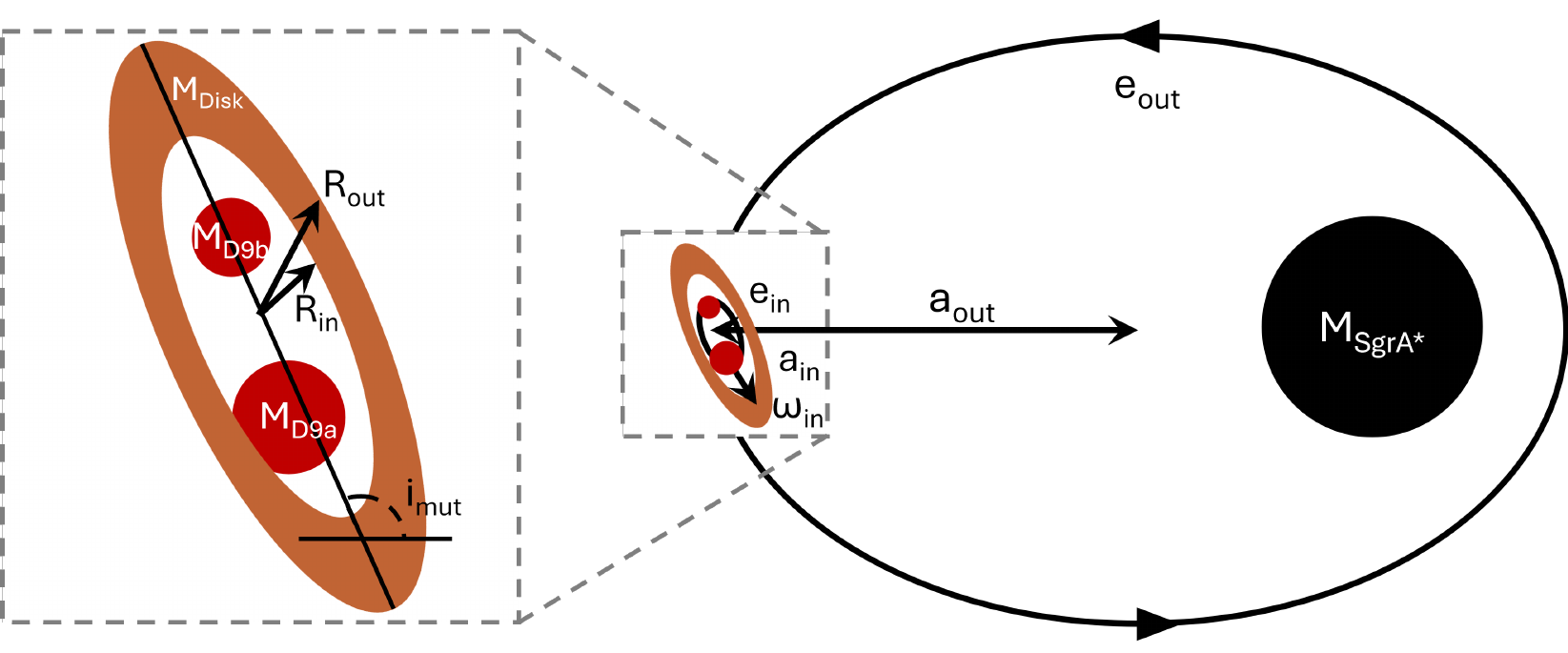}
    \caption{Sketch of the astrophysical setup. A stellar binary (D9) with circumbinary disk is orbiting a SMBH (Sgr A*). Masses ($M$), eccentricities ($e$), and semimajor axes ($a$) are indicated for the inner binary orbit and the outer orbit around the SMBH. The mutual inclination between these orbits is $i_{\rm mut}$ and the argument of periapsis of the inner orbit is $\omega_{\rm in}$.}
    \label{fig:setup}
\end{figure}

A sketch of the astrophysical setup of D9 is shown in \ref{fig:setup}. The orbital parameters and masses we used for our simulations are shown in Table \ref{tab:parameters}. 

\begin{table}[!h]
    \centering
    \caption{Initial conditions from \citet{Peissker_2024} for the simulation with the astrophysical setup as shown in Fig. \ref{fig:setup}.}
    \begin{tabular}{c|c}
    \hline
    \hline
        $e_{\rm in}$ & 0.45 \\
        $a_{\rm in}$ & 1.59 au \\
        $\omega_{\rm in}$ & $311.75^\circ$ \\
        $e_{\rm out}$ & 0.32 \\
        $a_{\rm out}$ & 44 mpc \\
        $i_{\rm mut}$ & $102.55^\circ$ \\
        $R_{\rm in}$ & 8.32 au \\
        $R_{\rm out}$ & 11.35 au \\
        $M_{\rm SgrA^*}$ & 4.297$\cdot10^6$ M$_{\odot}$ \\
        $M_{\rm D9a}$ & 2.8 M$_{\odot}$ \\
        $M_{\rm D9b}$ & 0.73 M$_{\odot}$\\
        $M_{\rm disk}$ & $1.6\cdot10^{-6}$ M$_{\odot}$\\
        \hline
    \end{tabular}
    
    \label{tab:parameters}
\end{table}

We modeled multiple spatial extents of the disk, with $R_{\rm in}$ and
$R_{\rm out}$ varying between a minimum inner radius and maximum outer
radius, both set by stability criteria. We set the minimum inner
radius to 4.5 au, which follows from the stable orbit of a test mass
around a binary as described by \citet{Mardling_2001} under the
assumption that the disk is circular and coplanar with the binary. We
set the maximum outer radius to 13.4 au, which corresponds to
one third of the Hill radius \citep{Hill_1878} of the binary. All
radii in this work are measured with respect to the center-of-mass of
the binary, unless otherwise specified.

\begin{figure*}
    \centering
    \includegraphics[width=\linewidth]{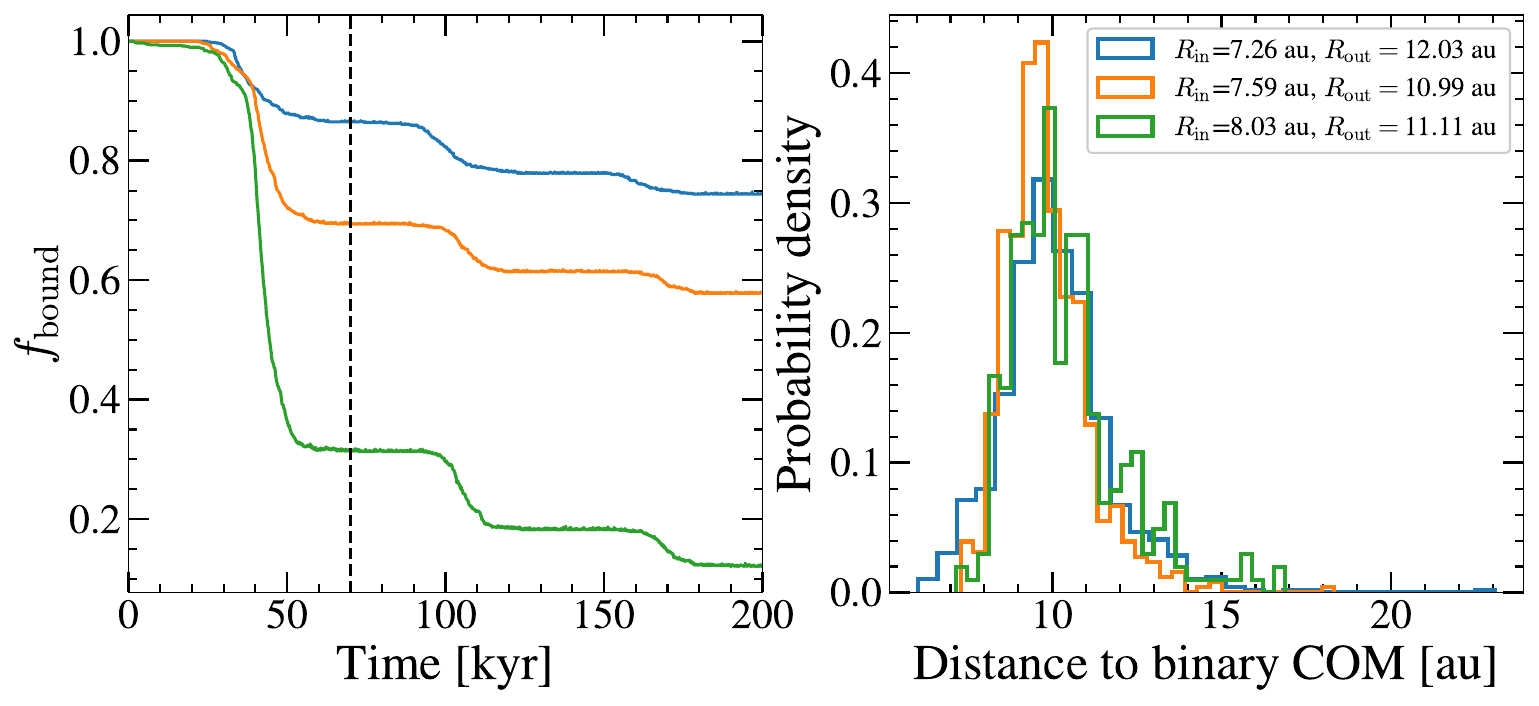}
    \caption{ \textit{Left:} Fraction of bound disk particles
      ($f_{\rm bound}$) is plotted as a function of time.  Different
      colors correspond to different values of the initial inner
      radius ($R_{\rm in}$) and initial outer radius ($R_{\rm out}$)
      of the circumbinary disk. The dashed black line indicates 70
      kyr, where the bound fraction has settled to its first plateau.
      \textit{Right:} Radial distribution of disk particles at 70
      kyr as a function of the distance to the binary
      center-of-mass. The disks in these preliminary runs consist of
      $10^3$ SPH particles.  }
    \label{fig:histograms}
\end{figure*}

From these preliminary runs, we observed that the system consistently converges to a quasi-stable configuration, characterized by periodic bursts of disk mass loss after an initial period of rapid mass loss (left panel of Fig. \ref{fig:histograms}). We defined the bound particles to be those on an orbit around the binary with an eccentricity of $e < 1$. We observe that these disks have very similar radial mass distributions (right panel of Fig. \ref{fig:histograms}). Motivated by this finding, we set the initial disk radii for our final runs by the bounds of the 68\% confidence interval of this radial mass distribution. Therefore, we chose $R_{\rm in} = 8.32$ au and $R_{\rm out} = 11.35$ au. All other parameters were kept the same for the final runs.

The orbital parameters of D9 as observed in \citet{Peissker_2024} could have arisen from the evolution of the system that triggered an episode of an eccentric mass transfer, leading to the circumbinary disk. Crossing the Roche limit of the stars on an eccentric orbit is a natural result of the vZLK process \citep{Stephan_2016}. In some systems, eccentric mass transfer can even excite eccentricity \citep{Rocha_2025}, potentially explaining the observed inner binary eccentricity and the circumbinary disk. Additionally, in some cases, tides and spin can even lead to further expansion of the system after mass transfer \citep{Cheng_2019}. In this scenario, the circumbinary disk is a consequence of the evolution (i.e., the mass transfer) rather than the initial condition. Additionally, a wider initial separation for the inner orbit could have led to stronger vZLK oscillations, causing the system we are seeing now on the track to merge (or shrink) rather than be two detached point masses on a stable orbit around each other.

\section{Results}\label{sect:results}

\begin{figure*}
    \centering
    \includegraphics[width=\linewidth]{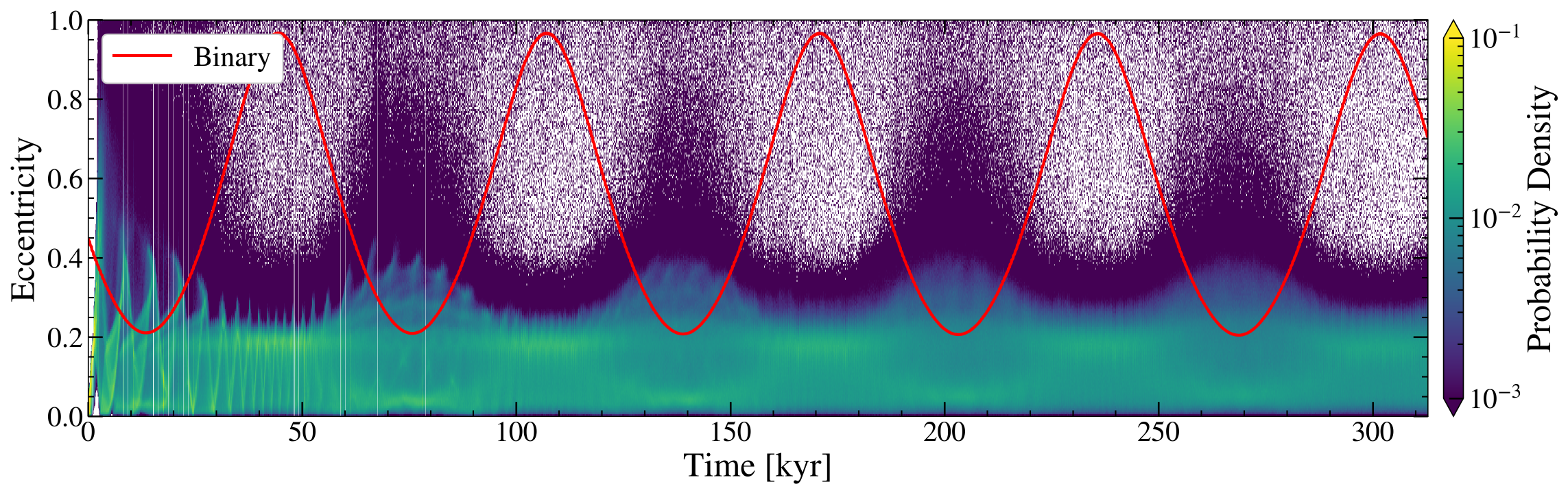}
    \includegraphics[width=\linewidth]{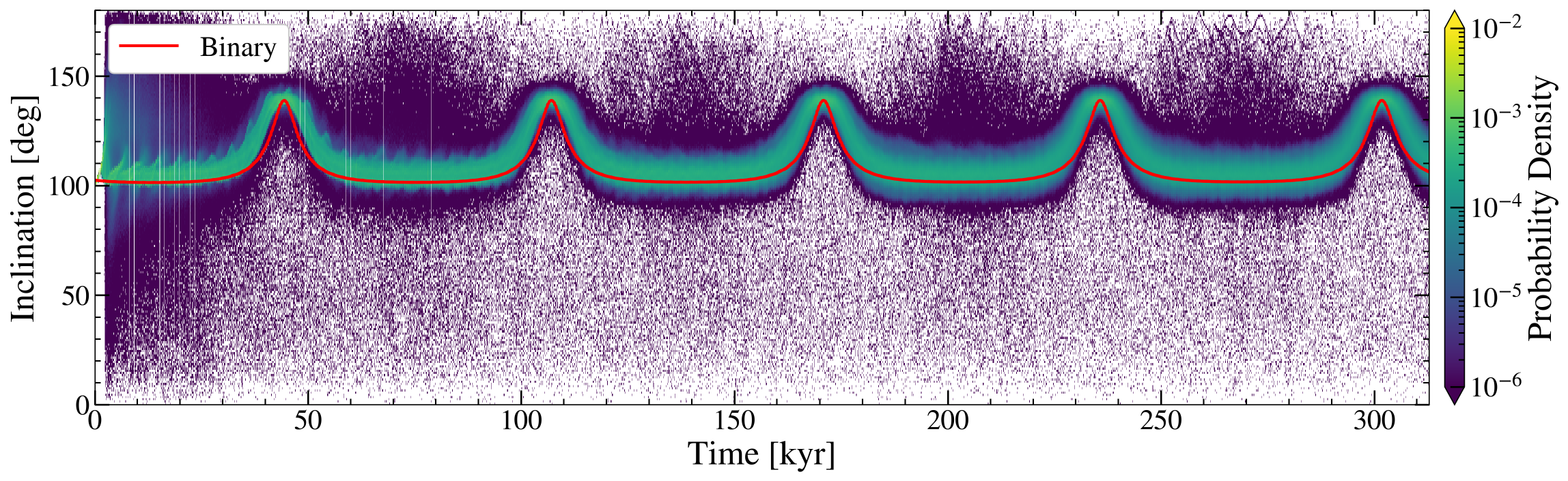}
    \includegraphics[width=\linewidth]{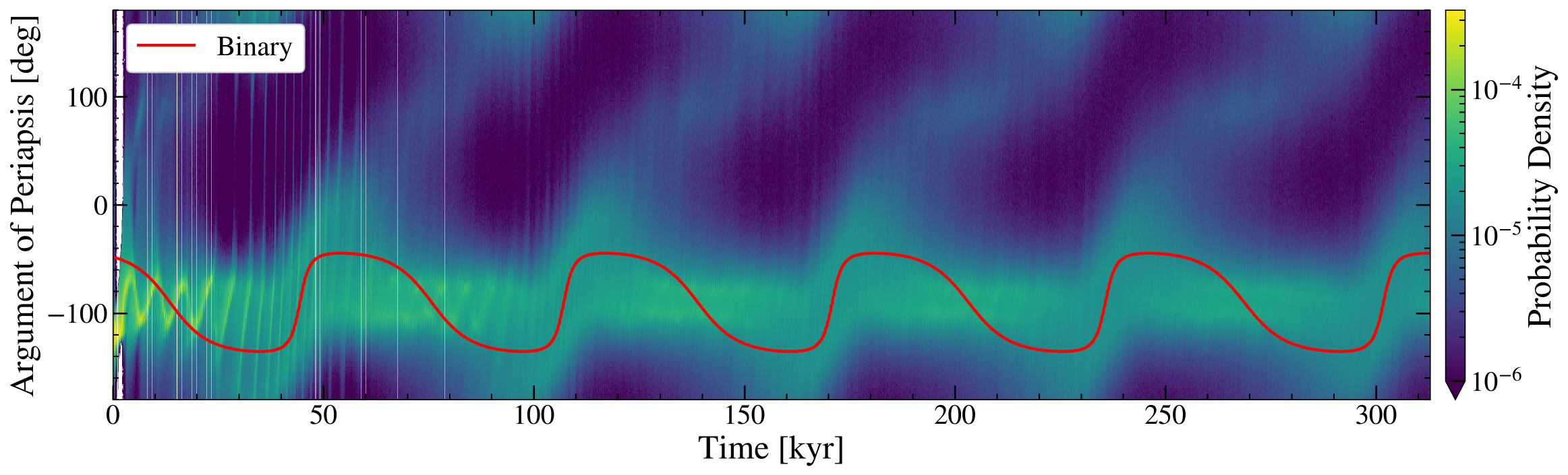}
    \includegraphics[width=\linewidth]{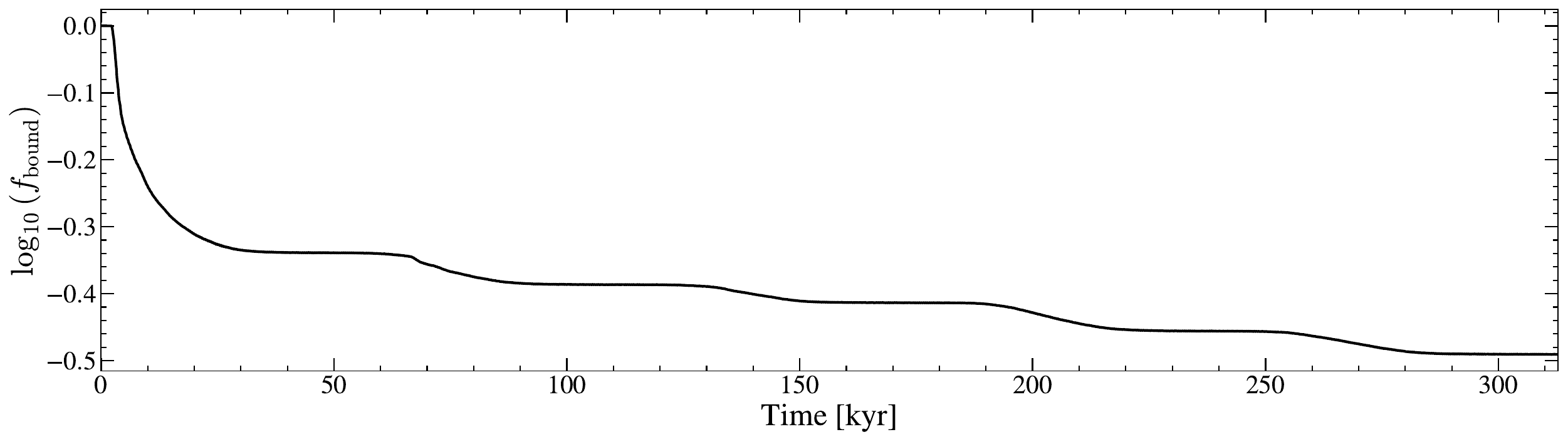}
    \caption{\textit{Top three panels:} Evolution of the distribution of the eccentricity, inclination, and the argument of periapsis of the bound ($e<1$) disk particles. The evolution of the binary orbital elements is also plotted (solid red lines). The orbital elements of the binary oscillate on a timescale of $T_{\rm vZLK} = 62.5$ kyr. \textit{Bottom:} Fraction of bound disk particles over time. After an initial period of rapid mass loss, the disk loses mass in periodic bursts on the vZLK timescale.}
    \label{fig:secular_evolution}
\end{figure*}

\subsection{Secular evolution of D9}
In the top panel of Fig. \ref{fig:secular_evolution}, we plot the time evolution of the eccentricity of the binary and disk components of D9. The eccentricity of the binary oscillates on a timescale of $T_{\rm vZLK} \approx$ 62.5 kyr. The eccentricity varies between $e_{\rm min} \approx$ 0.20 and $e_{\rm max} \approx$ 0.97. Using Eq. 20 in \citet{Hamers_2021}, we calculate a theoretical vZLK timescale of $T_{\rm vZLK} = 60.5$ kyr on which the eccentricity oscillates between $e_{\rm min} \approx 0.22$ and $e_{\rm max} \approx 0.97$, in good agreement with our simulated values. 

The eccentricity distribution of the disk particles shows the same periodicity on the vZLK timescale, though approximately in antiphase with the binary. There is also a shorter-period fluctuation visible with a periodicity of about 2 kyr when the binary has its highest eccentricity, and increasing to about 4 kyr when less eccentric. Such a quasi-periodicity could be an indication of Hopf bifurcation in the disc, which would indicate a chaotic response on the disc, as its quasi-periodicity is driven by the orbiting stars. But it could be a simulation artifact of the disk stabilizing due to the way it was initialized, because these minor fluctuations dampen over time, smoothing out after one vZLK cycle.
 
The second panel of Fig. \ref{fig:secular_evolution} shows the inclination over time. The peaks of the binary inclination coincide with the peaks of the binary eccentricity, which is in line with the expected behavior. We see the bulk of the disk traces the inclination variations of the binary.

The third panel of Fig. \ref{fig:secular_evolution} shows the argument of periapsis over time. The bulk of the disk again roughly follows the evolution of the argument of periapsis of the binary. We also see shorter-timescale oscillations, similar to the oscillations in the eccentricity evolution.

\subsection{vZLK-driven disk mass loss}
The bottom panel of Fig. \ref{fig:secular_evolution} shows the number of bound disk particles over time. As seen in our preliminary runs (see Sect. \ref{sect:methods}), the disk undergoes an initial period of rapid mass loss, after which it settles in a quasi-stable configuration in which it exhibits periodic bursts of mass loss on a timescale similar to $T_{\rm vZLK}$. This suggests that the mass loss is driven by the vZLK mechanism. The bursts coincide with the eccentricity distribution of the disk extending to higher values, as is seen in the top panel of Fig. \ref{fig:secular_evolution}.

\section{Discussion}\label{sect:discussion}

\subsection{The mass evolution of the circumbinary disk}
We have shown that a disk orbiting the D9 binary periodically loses mass on a timescale corresponding to $T_{\rm  vZLK} = 62.5\,$kyr. Each secular cycle, the disk loses 7\% $\pm$ 2\% of its mass. If we assume that mass loss persisted throughout the estimated binary system lifetime of $2.7^{+1.9}_{-0.3}\,$Myr, we arrive at an initial disk mass of $M_{\rm disk} = 1.4 \cdot 10^{-4}$ M$_{\odot}$. Since we do not account for other sources of mass loss, such as stellar winds, this is a lower bound on the initial disk mass.

To verify our results, we ran two additional simulations using the same parameters as in Table \ref{tab:parameters}, except for the disk mass, which we set to $M_{\rm disk} = 1.4\cdot 10^{-4}$ M$_{\odot}$ and $M_{\rm disk} = 1.5\cdot 10^{-5}$ M$_{\odot}$. The former is the initial mass estimate based on preliminary results (as shown in Fig. \ref{fig:histograms}) and the latter the geometric mean of this estimated initial mass and the observed disk mass. Changing the initial disk mass does not affect our results.

\subsection{Lifetime of D9}
Assuming that the disk loses $7$\% of its mass every $T_{\rm  vZLK} = 62.5\,$kyr, the disk retains 1\% of its current mass after $\sim$4 Myr, at which point detection is unlikely. D9 will then be  $\sim$6.7 Myr old. Therefore, if D9 is born with a circumbinary disk, we estimate the time window within which the disk of D9 is detectable to be $\sim$6.7 Myr. This is an upper limit, because we disregard other sources of mass loss, such as photo-evaporation \citep{Alexander_2012} or close gravitational encounters with other objects in the Galactic center. 
The absence of Br$\gamma$ emission around other stars in the S cluster can be explained by vZLK-driven mass loss of a circumbinary disk, because the average age of massive early-type S cluster members is $\sim$4-6 Myr \citep{Lu2013, Habibi2017}. This is similar to the age of D9 by the time its disk has become undetectable.


\subsection{The potential future merger of the binary D9}
The vZLK effect can drive binaries to highly eccentric orbits, possibly decreasing the binary separation to the regime of Roche-lobe overflow and subsequent merger. \citet{Peissker_2024} argue that, if the vZLK effect is strong enough to drive D9 to Roche-lobe overflow within a vZLK timescale, this is expected in the next vZLK timescale. They hypothesize that the population of G objects consists of pre-merger binaries (such as D9) and post-merger products (D9 merger products; \citealt{Peissker2024}).

D9 has undergone many vZLK cycles during its estimated lifetime. However, since we model the stars in D9 as point masses and neglect stellar evolution, we do not allow a merger of the binary in our simulations. The stars in our simulation never come closer to each other than $(1 - e_{\rm max}) a_{\rm in} \gtrsim 4 R_{\star}$. Tidal effects may have played a role in the past evolution of D9. Such tidal evolution would naturally lead to the circularization of the binary. The current relatively high eccentricity suggests that tides are not important for this system on the timescale of its lifetime.

We tested the effects of tides in D9 by simulating the evolution of the
system, using a secular approach including the tidal evolution of the
stars \citep{Hamers_2016}. We did this using the
\texttt{SecularMultiple} code through the \texttt{AMUSE} interface
\citep{Portegies_Zwart_2013}. For both stellar radii, we
conservatively\footnote{Evolving the two stars for $2.7$ Myr with
stellar evolution code \texttt{SeBa} \citep{spz_1996,Toonen_2012}, we
find a radius of 1.91 $R_{\odot}$ for the primary and 0.68 $R_{\odot}$
for the secondary.} adopted $2 \, R_{\odot}$. We find no appreciable
change in the secular evolution of the orbital elements of the inner
binary over several vZLK timescales, compared to our direct N-body
simulations. We conclude that the vZLK effect will not drive the
binary D9 to merger.

\section{Summary and conclusions}\label{sect:conclusions}

We simulated the recently discovered S-star binary in the Galactic center \citep{Peissker_2024}, with its circumbinary disk. Our calculations were performed using \texttt{AMUSE} to couple a smoothed particles hydrodynamics code with a direct N-body code. We assumed that the binary D9 is on a stable orbit; however, it is also possible that the observed orbital parameters and disk are the result of an earlier episode of eccentric mass transfer rather than these initial conditions. Based on the stable orbit scenario, we conclude that:
\begin{enumerate}
    \item The binary D9 experiences vZLK oscillations with a period of $T_{\rm vZLK} \approx$ 62.5 kyr, during which the eccentricity varies between $e_{\rm min} \approx$ 0.20 and $e_{\rm max} \approx$ 0.97. This is consistent with theory.
    \item The orbital elements of the disk particles show a periodicity on the vZLK timescale. The eccentricity distribution of circumbinary disk material is widest when the binary is at its minimum eccentricity. The distributions of inclination, argument of periapsis, and longitude of the ascending node trace the secular evolution of the binary.
    \item The disk loses approximately 7\% $\pm$ 2\% of its mass in periodic bursts on the vZLK timescale. These bursts coincide with the broadening of the disk material eccentricity distribution.
    \item Extrapolating the periodic mass loss over the lifetime of D9 and assuming that this is the only source of disk mass loss, we find a lower limit to the initial disk mass of 1.4$\cdot10^{-4}$ M$_{\odot}$.
    \item Extrapolating the vZLK-driven mass loss to the future, we find that the disk will have 1\% of its current mass after $\sim$ 4 Myr. The age of D9 will then be $\sim$6.7 Myr.

    \item The absence of Br$\gamma$ emission around other stars in the
      S cluster can be explained by vZLK-driven mass loss of a
      circumbinary disk. However, it is also possible that a
      circumbinary disk is never acquired, or that there are
      additional processes destroying it.  At the same time, if S-star
      clusters are a merger product \cite[as discussed in
      ][]{Stephan_2016,2019ApJ...878...58S} then their circumbinary
      disk would dissipate on a short timescale, as our calculations
      demonstrate.
\end{enumerate}

\begin{acknowledgements} 
The authors thank Erwan Hochart and Gijs Vermariën for useful
discussions. The authors thank the referee for their constructive
feedback. This work was performed using the compute resources from the
Academic Leiden Interdisciplinary Cluster Environment (ALICE) provided
by Leiden University. This work resulted from the master's course
Simulation and Modeling in Astrophysics at Leiden Observatory.
\end{acknowledgements}

\bibliographystyle{aa} 

\end{document}